\documentstyle[12pt]{article}
\def\3{\ss}
 \renewcommand{\arraystretch}{2.5}
 \newcommand{\HG}{H(4)}

\newcommand{\half}{\mbox{\small $\frac{1}{2}$}}

\newcommand{\Dd}[1]{\mbox{
  \parbox[b]{0cm}{$D$}\raisebox{1.7ex}{$\leftrightarrow$}$_{\!#1}$}}

\newcommand{\symm}{
                     \unitlength0.3cm
                     \begin{picture}(4.5,1.5)
                     \put(0,0){\line(0,1){1}}
                     \put(1,0){\line(0,1){1}}
                     \put(2,0){\line(0,1){1}}
                     \put(3,0){\line(0,1){1}}
                     \put(4,0){\line(0,1){1}}
                     \put(0,0){\line(1,0){4}}
                     \put(0,1){\line(1,0){4}}
                     \end{picture}               }
\newcommand{\liha}{
                     \unitlength0.3cm
                     \begin{picture}(3.5,2.5)
                     \put(0,0){\line(0,1){2}}
                     \put(1,0){\line(0,1){2}}
                     \put(2,1){\line(0,1){1}}
                     \put(3,1){\line(0,1){1}}
                     \put(0,0){\line(1,0){1}}
                     \put(0,1){\line(1,0){3}}
                     \put(0,2){\line(1,0){3}}
                     \end{picture}               }
\newcommand{\kast}{
                     \unitlength0.3cm
                     \begin{picture}(2.5,2.5)
                     \put(0,0){\line(0,1){2}}
                     \put(1,0){\line(0,1){2}}
                     \put(2,0){\line(0,1){2}}
                     \put(0,0){\line(1,0){2}}
                     \put(0,1){\line(1,0){2}}
                     \put(0,2){\line(1,0){2}}
                     \end{picture}              }
\newcommand{\stha}{
                     \unitlength0.3cm
                     \begin{picture}(2.5,3.5)
                     \put(0,0){\line(0,1){3}}
                     \put(1,0){\line(0,1){3}}
                     \put(2,2){\line(0,1){1}}
                     \put(0,0){\line(1,0){1}}
                     \put(0,1){\line(1,0){1}}
                     \put(0,2){\line(1,0){2}}
                     \put(0,3){\line(1,0){2}}
                     \end{picture}               }
\newcommand{\anti}{
                     \unitlength0.3cm
                     \begin{picture}(1.5,4.5)
                     \put(0,0){\line(0,1){4}}
                     \put(1,0){\line(0,1){4}}
                     \put(0,0){\line(1,0){1}}
                     \put(0,1){\line(1,0){1}}
                     \put(0,2){\line(1,0){1}}
                     \put(0,3){\line(1,0){1}}
                     \put(0,4){\line(1,0){1}}
                     \end{picture}               }
\begin{document}
%
\makeatletter
\@addtoreset{equation}{section}
\makeatother
\renewcommand{\theequation}{\thesection.\arabic{equation}}
\begin{titlepage}
\thispagestyle{empty}

\begin{minipage}{13.25cm}

 \vspace{-2cm}

    February, 1996   \hfill
 \begin{minipage}[t]{2.7cm}
 DESY 96-031 \\
 HLRZ 96-11 \\
 HUB-EP-96/4
 \end{minipage}
\begin{LARGE}
  \begin{center}
      {\bf{Lattice Operators for Moments of the Structure Functions
           and their Transformation under the Hypercubic Group}}
  \end{center}
\end{LARGE}
\begin{center}
        M. G\"ockeler$^{1,2}$,
        R. Horsley$^3$, E.-M. Ilgenfritz$^3$, H. Perlt$^4$,
        P. Rakow$^1$, G.~Schierholz$^{5,1}$ and A. Schiller$^4$\\[2em]
        $^1$ HLRZ, c/o Forschungszentrum J{\"u}lich,
                    D-52425 J{\"u}lich,
                    Germany\\[0.5em]
        $^2$ Institut f\"ur Theoretische Physik,
                    RWTH Aachen, \\
                    D-52056 Aachen, Germany\\[0.5em]
        $^3$ Institut f\"ur Physik, Humboldt-Universit\"at,
                    D-10115 Berlin, Germany\\[0.5em]
        $^4$ Institut f\"ur Theoretische Physik,
                    Universit\"at Leipzig,\\
                    D-04109 Leipzig, Germany\\[0.5em]
        $^5$ Deutsches Elektronen-Synchrotron DESY,
             D-22603 Hamburg, Germany
\end{center}

\end{minipage}

 \vspace{3cm}

 \centerline{\bf Abstract}
 \begin{quote}
 For lattice operators that are relevant to the calculation of
 moments of nucleon structure functions we investigate the
 transformation properties under the hypercubic group.
 We give explicit bases of irreducible subspaces for tensors
 of rank $\leq 4$.
 \end{quote}

\end{titlepage}

\section{Introduction}

The calculation of the moments of hadronic structure functions by
Monte Carlo simulations of lattice QCD requires the study of matrix
elements of composite operators appearing in the operator product
expansion of the appropriate currents \cite{mart,capi,roma,letter}.
In the continuum, these
operators are classified according to their behaviour under
Lorentz transformations and charge conjugation. In particular, the
leading twist (=2) operators are totally symmetric and traceless
Lorentz tensors. Due to their transformation properties they
cannot mix with lower-dimensional operators under renormalization.

When QCD is put on the lattice, the necessary analytic continuation
from Minkowski to Euclidean space replaces the Lorentz group by the
orthogonal group $O(4)$, which by the discretization of space-time
is further reduced to the hypercubic group $\HG \subset O(4)$.
Hence the lattice operators have to be classified according to
$\HG$ and one should only work with operators which transform
irreducibly under $\HG$.
Since $\HG$ is only a finite group, the restrictions imposed
by symmetry are less stringent than in the continuum and the
possibilities for mixing increase. A control over the mixing
problem is however vital for a successful lattice calculation.
In particular, mixing with lower-dimensional operators presents
special difficulties and is to be avoided whenever possible.
Therefore a computation of structure functions on the lattice
requires a study of the $\HG$-transformation properties of the
relevant lattice operators.

In the present paper we give explicit bases for $\HG$-irreducible
subspaces of operators up to spin 4.
We consider the following types of operators:
\begin{equation}
\label{op0}
{\cal O}^0_{\mu_1 \mu_2 \ldots \mu_n} :=
 \bar{\psi} \gamma_{\mu_1} \Dd{\mu_2} \Dd{\mu_3} \cdots
 \Dd{\mu_n} \psi\,,
\end{equation}
\begin{equation}
\label{op5}
{\cal O}^5_{\mu_1 \mu_2 \ldots \mu_n} :=
 \bar{\psi} \gamma_{\mu_1} \gamma_5
 \Dd{\mu_2} \Dd{\mu_3} \cdots \Dd{\mu_n} \psi\,,
\end{equation}
\begin{equation}
{\cal O}^g_{\mu_1 \mu_2 \ldots \mu_n} :=
 \mbox{tr} \sum^4_{\alpha = 1}
 G_{\mu_1 \alpha} D_{\mu_2} \cdots D_{\mu_{n-1}} G_{\mu_n \alpha}\,.
\end{equation}
Here $\psi$ is a Wilson fermion field,
$D_\mu$ denotes the lattice covariant derivative
and $\!\!\Dd{\mu} =\mbox{
\parbox[b]{0cm}{$D_\mu$}\raisebox{1.7ex}{$\rightarrow$}}  -
\mbox{\parbox[b]{0cm}{$D_\mu$}\raisebox{1.7ex}{$\leftarrow$}} \,\,$
(for the proper lattice definition of expressions like
\parbox[b]{0cm}{$D_\mu$}\raisebox{1.7ex}{$\rightarrow$}
\parbox[b]{0cm}{$D_\nu$}\raisebox{1.7ex}{$\leftarrow \,\,$}
see \cite{roma}).
The (clover) lattice version $  G_{\mu \nu}$
of the gluon field strength is given by
\begin{equation} \begin{array}{r} \displaystyle
G_{\mu \nu} (x) = \frac{\mbox{i}}{8} \sum_{s, s^\prime = \pm 1}
  s s^\prime U(x,x+s \hat{\nu}) U(x+s \hat{\nu}, x+s \hat{\nu}
     +s^\prime \hat{\mu})  \qquad \\ \displaystyle
{} \times U(x+s \hat{\nu} +s^\prime \hat{\mu},
   x +s^\prime \hat{\mu}) U(x+s^\prime \hat{\mu},x) + \mbox{h.c.} \,,
\end{array}
\end{equation}
where we have defined the parallel transporter $U(x,y)$ for
nearest-neighbour lattice points $x,y$ as
\begin{equation}
\renewcommand{\arraystretch}{1.5}
 U(x,y) = \left\{   \begin{array}{ll}
    U(x,\mu)              &  \mbox{if} \;\; y = x + \hat{\mu} \,, \\
    U(x-\hat{\mu},\mu)^+  &  \mbox{if} \;\; y = x - \hat{\mu} \,.
   \end{array} \right.
\end{equation}

\section{Transformation Properties}

Charge conjugation operates on the fermion fields $\psi (x)$,
$\bar{\psi} (x)$ and on the lattice gauge field $U(x,\mu)$
according to
\begin{eqnarray}
 \psi (x) & \to & C \bar{\psi} (x) ^T  \,, \\
 \bar{\psi} (x) & \to & - \psi(x) ^T C^{-1} \,, \\
 U(x,\mu)       & \to & U(x,\mu)^*
\end{eqnarray}
with the charge conjugation matrix $C$ satisfying
\begin{equation}
 C \gamma_\mu^T C^{-1} = - \gamma_\mu
\end{equation}
and we get
\begin{equation} \label{chacon}
{\cal O}^0_{\mu_1 \mu_2 \ldots \mu_n} \to
(-1)^n {\cal O}^0_{\mu_1 \mu_n \mu_{n-1} \ldots \mu_2 }  \,,
\end{equation}
\begin{equation}
{\cal O}^5_{\mu_1 \mu_2 \ldots \mu_n} \to
(-1)^{n-1} {\cal O}^5_{\mu_1 \mu_n \mu_{n-1} \ldots \mu_2 }  \,,
\end{equation}
\begin{equation}
{\cal O}^g_{\mu_1 \mu_2 \ldots \mu_n} \to
{\cal O}^g_{\mu_1 \mu_2 \ldots \mu_n} \,.
\end{equation}
Note that replacing $\Dd{\mu}$ by
\parbox[b]{0cm}{$D_\mu$}\raisebox{1.7ex}{$\rightarrow \,$} or
\parbox[b]{0cm}{$D_\mu$}\raisebox{1.7ex}{$\leftarrow \,$}
in (\ref{op0}), (\ref{op5})
leads to operators that have no simple behaviour under charge
conjugation.

In the study of the transformation properties under $\HG$
we use the notation of ref.\cite{bonn} and consider
\begin{equation}
\HG = \left\{ (a,\pi) | a \in Z_2^4 \,,\, \pi \in S_4 \right\} \,.
\end{equation}
So $\pi$ is a permutation of four elements and
$a_\mu \in\{0,1\}$.
For elements of the symmetric group $S_n$
we shall use the cycle notation (see, e.g., \cite{miller}).
In terms of an orthonormal basis $v_1,\ldots,v_4$ of
a four-dimensional vector space the defining representation $T$
of $\HG$ is given by:
\begin{equation}
T(a,\pi) v_\mu = (-1)^{a_\mu} v_{\pi(\mu)}  \,.
\end{equation}
The corresponding orthogonal $4 \times 4$ matrices are
\begin{equation}
T(a,\pi)_{\mu \nu} = (-1)^{a_\nu} \delta_{\mu,\pi(\nu)} \,.
\end{equation}

Note that $\HG$ is generated by the three elements
\begin{eqnarray}
\label{genalpha}
\alpha & = & \bigl( (0,0,0,0),(12) \bigr) \,,  \\
\beta  & = & \bigl( (0,0,0,0),(2341) \bigr) \,,  \\
\label{gengamma}
\gamma & = & \bigl( (1,0,0,0),\mbox{id} \bigr) \,.
\end{eqnarray}
The corresponding matrices are given by
\renewcommand{\arraystretch}{1.5}
\begin{eqnarray}
T(\alpha) & = & \left (
   \begin{array}{cccc}
       0 & 1 & 0 & 0  \\
       1 & 0 & 0 & 0  \\
       0 & 0 & 1 & 0  \\
       0 & 0 & 0 & 1    \end{array}   \right ) \,, \\
T(\beta ) & = & \left (
   \begin{array}{cccc}
       0 & 0 & 0 & 1  \\
       1 & 0 & 0 & 0  \\
       0 & 1 & 0 & 0  \\
       0 & 0 & 1 & 0    \end{array}   \right ) \,, \\
T(\gamma) & = & \left (
   \begin{array}{cccc}
      -1 & 0 & 0 & 0  \\
       0 & 1 & 0 & 0  \\
       0 & 0 & 1 & 0  \\
       0 & 0 & 0 & 1    \end{array}   \right ) \,.
\end{eqnarray}
\renewcommand{\arraystretch}{2.5}

On our operators the element $(a,\pi) \in \HG$ acts as follows:
\begin{equation} \label{trans0}
{\cal O}^0_{\mu_1 \mu_2 \ldots \mu_n} \to
(-1)^{a_{\mu_1}+a_{\mu_2}+\cdots +a_{\mu_n}}
{\cal O}^0_{\pi(\mu_1) \pi(\mu_2) \ldots \pi(\mu_n)}  \,,
\end{equation}
\begin{equation}
{\cal O}^g_{\mu_1 \mu_2 \ldots \mu_n} \to
(-1)^{a_{\mu_1}+a_{\mu_2}+\cdots +a_{\mu_n}}
{\cal O}^g_{\pi(\mu_1) \pi(\mu_2) \ldots \pi(\mu_n)}  \,,
\end{equation}
\begin{equation} \label{trans5}
{\cal O}^5_{\mu_1 \mu_2 \ldots \mu_n} \to
(-1)^{a_{\mu_1}+a_{\mu_2}+\cdots +a_{\mu_n}} \det ( T(a,\pi))
{\cal O}^5_{\pi(\mu_1) \pi(\mu_2) \ldots \pi(\mu_n)}  \,.
\end{equation}
So both ${\cal O}^0$ and ${\cal O}^g$ transform according to
the $n$-th tensor power of $T$.
The operators
${\cal O}^5$, on the other hand, have one factor of $T$ replaced
by $T \cdot \det T$. Therefore these tensor
products of $\HG$ representations have to be decomposed
into irreducible components. In doing so we consider the operators
(\ref{op0}) ((\ref{op5}), respectively) as forming an orthonormal
basis of the representation space. The bases for $\HG$-irreducible
subspaces to be given below will also be chosen orthonormal.
We shall present
our results only for ${\cal O}^0$, the formulas for ${\cal O}^g$
are identical, except for the $C$-parity.

\section{Irreducible Representations and Decomposition}

The 20 (inequivalent) irreducible representations of $\HG$ are
denoted by $\tau^{(l)}_k$, where $l$ is the dimension of the
representation and $k=1,2,\ldots$ distinguishes inequivalent
representations of the same dimension. There are four one-dimensional
representations, two of dimension two, four of dimension three,
four and six, and two of dimension eight. The
defining representation $T$ will be labeled as $\tau^{(4)}_1$ and
the representation $T \cdot \det T$ as $\tau^{(4)}_4$.
For the reader's convenience, we give in Table 1 the correspondence
with the notation of ref.\cite{mand}.
\begin{table}[t]
 \renewcommand{\arraystretch}{1.5}
\hspace*{2.5cm}
\begin{minipage}[t]{6.0cm}
\begin{tabular}{|c|c|} \hline
\cite{bonn}      & \cite{mand}                             \\ \hline
$ \tau^{(1)}_1 $ & $ I^{(+)} $                             \\
\raisebox{0.6cm}{$\tau^{(1)}_2 $}
                 &   \unitlength0.3cm
                     \begin{picture}(2.5,4.5)
                     \put(0,0){\line(0,1){4}}
                     \put(1,0){\line(0,1){4}}
                     \put(0,0){\line(1,0){1}}
                     \put(0,1){\line(1,0){1}}
                     \put(0,2){\line(1,0){1}}
                     \put(0,3){\line(1,0){1}}
                     \put(0,4){\line(1,0){1}}
                     \put(2.2,4.2){\makebox(0,0){$^{(-)}$}}
                     \end{picture}                         \\
\raisebox{0.6cm}{$\tau^{(1)}_3 $}
                 &   \unitlength0.3cm
                     \begin{picture}(2.5,4.5)
                     \put(0,0){\line(0,1){4}}
                     \put(1,0){\line(0,1){4}}
                     \put(0,0){\line(1,0){1}}
                     \put(0,1){\line(1,0){1}}
                     \put(0,2){\line(1,0){1}}
                     \put(0,3){\line(1,0){1}}
                     \put(0,4){\line(1,0){1}}
                     \put(2.2,4.2){\makebox(0,0){$^{(+)}$}}
                     \end{picture}                         \\
$ \tau^{(1)}_4 $ & $ I^{(-)} $                             \\
\raisebox{0.2cm}{$\tau^{(2)}_1 $}
                 &   \unitlength0.3cm
                     \begin{picture}(3.5,2.5)
                     \put(0,0){\line(0,1){2}}
                     \put(1,0){\line(0,1){2}}
                     \put(2,0){\line(0,1){2}}
                     \put(0,0){\line(1,0){2}}
                     \put(0,1){\line(1,0){2}}
                     \put(0,2){\line(1,0){2}}
                     \put(3.2,2.2){\makebox(0,0){$^{(+)}$}}
                     \end{picture}                         \\
\raisebox{0.2cm}{$\tau^{(2)}_2 $}
                 &   \unitlength0.3cm
                     \begin{picture}(3.5,2.5)
                     \put(0,0){\line(0,1){2}}
                     \put(1,0){\line(0,1){2}}
                     \put(2,0){\line(0,1){2}}
                     \put(0,0){\line(1,0){2}}
                     \put(0,1){\line(1,0){2}}
                     \put(0,2){\line(1,0){2}}
                     \put(3.2,2.2){\makebox(0,0){$^{(-)}$}}
                     \end{picture}                         \\
\raisebox{0.2cm}{$\tau^{(3)}_1 $}
                 &   \unitlength0.3cm
                     \begin{picture}(4.5,2.5)
                     \put(0,0){\line(0,1){2}}
                     \put(1,0){\line(0,1){2}}
                     \put(2,1){\line(0,1){1}}
                     \put(3,1){\line(0,1){1}}
                     \put(0,0){\line(1,0){1}}
                     \put(0,1){\line(1,0){3}}
                     \put(0,2){\line(1,0){3}}
                     \put(4.2,2.2){\makebox(0,0){$^{(+)}$}}
                     \end{picture}                         \\
\raisebox{0.35cm}{$\tau^{(3)}_2 $}
                 &   \unitlength0.3cm
                     \begin{picture}(4.5,3.5)
                     \put(0,0){\line(0,1){3}}
                     \put(1,0){\line(0,1){3}}
                     \put(2,2){\line(0,1){1}}
                     \put(0,0){\line(1,0){1}}
                     \put(0,1){\line(1,0){1}}
                     \put(0,2){\line(1,0){2}}
                     \put(0,3){\line(1,0){2}}
                     \put(3.2,3.2){\makebox(0,0){$^{(-)}$}}
                     \end{picture}                         \\
\raisebox{0.35cm}{$\tau^{(3)}_3 $}
                 &   \unitlength0.3cm
                     \begin{picture}(3.5,3.5)
                     \put(0,0){\line(0,1){3}}
                     \put(1,0){\line(0,1){3}}
                     \put(2,2){\line(0,1){1}}
                     \put(0,0){\line(1,0){1}}
                     \put(0,1){\line(1,0){1}}
                     \put(0,2){\line(1,0){2}}
                     \put(0,3){\line(1,0){2}}
                     \put(3.2,3.2){\makebox(0,0){$^{(+)}$}}
                     \end{picture}                         \\
\raisebox{0.2cm}{$\tau^{(3)}_4 $}
                 &   \unitlength0.3cm
                     \begin{picture}(4.5,2.5)
                     \put(0,0){\line(0,1){2}}
                     \put(1,0){\line(0,1){2}}
                     \put(2,1){\line(0,1){1}}
                     \put(3,1){\line(0,1){1}}
                     \put(0,0){\line(1,0){1}}
                     \put(0,1){\line(1,0){3}}
                     \put(0,2){\line(1,0){3}}
                     \put(4.2,2.2){\makebox(0,0){$^{(-)}$}}
                     \end{picture}                         \\ \hline
\end{tabular}
\end{minipage}
\begin{minipage}[t]{6.0cm}
\begin{tabular}{|c|c|} \hline
\cite{bonn}      & \cite{mand}                             \\ \hline
$ \tau^{(4)}_1 $ & $ (\half,\half)^{(+)} $                 \\
$ \tau^{(4)}_2 $ & $ \overline{(\half,\half)}^{(-)} $      \\
$ \tau^{(4)}_3 $ & $ \overline{(\half,\half)}^{(+)} $      \\
$ \tau^{(4)}_4 $ & $ (\half,\half)^{(-)} $                 \\
$ \tau^{(6)}_1 $ & $ (10) \oplus (01) $                    \\
$ \tau^{(6)}_2 $ & $ (\overline{10}) \oplus (\overline{01}) $ \\
$ \tau^{(6)}_3 $ & $ 6^{(+)} $                             \\
$ \tau^{(6)}_4 $ & $ 6^{(-)} $                             \\
$ \tau^{(8)}_1 $ & $ 8^{(+)} $                             \\
$ \tau^{(8)}_2 $ & $ 8^{(-)} $                             \\ \hline
\end{tabular}
\end{minipage}
 \caption{  Irreducible representations of $\HG$. }
\end{table}
In the following, the symbol ``=''  used for representations
only indicates equivalence and not actual identity of the
representations.

For the tensor products of $\tau^{(4)}_1$ with
itself we have the following decompositions in the cases
$n=2,3,4$:
\begin{equation}
\renewcommand{\arraystretch}{1.5}
\begin{array}{l} \displaystyle
\tau^{(4)}_1 \otimes \tau^{(4)}_1 =
\tau^{(1)}_1 \oplus \tau^{(3)}_1 \oplus
\tau^{(6)}_1 \oplus \tau^{(6)}_3 \,,
 \\ \displaystyle
\tau^{(4)}_1 \otimes \tau^{(4)}_1  \otimes \tau^{(4)}_1 =
4 \tau^{(4)}_1 \oplus \tau^{(4)}_2 \oplus
\tau^{(4)}_4 \oplus 3 \tau^{(8)}_1
\oplus 2 \tau^{(8)}_2 \,,
 \\ \displaystyle
\tau^{(4)}_1 \otimes \tau^{(4)}_1  \otimes \tau^{(4)}_1
                                     \otimes \tau^{(4)}_1
 \\ \displaystyle \qquad  { } =
4 \tau^{(1)}_1 \oplus \tau^{(1)}_2 \oplus \tau^{(1)}_4
\oplus 3 \tau^{(2)}_1 \oplus 2 \tau^{(2)}_2 \oplus 7 \tau^{(3)}_1
\oplus 3 \tau^{(3)}_2 \oplus 3 \tau^{(3)}_3 \oplus 3 \tau^{(3)}_4
 \\ \displaystyle \hphantom{\qquad  { } = } { }
\oplus 10 \tau^{(6)}_1 \oplus 6 \tau^{(6)}_2 \oplus 10 \tau^{(6)}_3
\oplus 6 \tau^{(6)}_4 \,.
\end{array}
\end{equation}

For $n=2,3$ and the totally symmetric tensors of rank 4 one can
construct bases for irreducible subspaces simply by inspection. For
the not totally symmetric rank-4 tensors we have to proceed more
systematically as will be explained below.
In the end one obtains a
list of operators which have definite charge conjugation
parity $C$ and transform irreducibly under $\HG$. Furthermore,
they can be classified as being traceless (i.e.\ all possible
traces vanish) or having nonvanishing trace.
They also belong
to definite irreducible representations of the symmetric group $S_n$
acting according to
\begin{equation}
\label{symmact}
P(p^{-1}) {\cal O}^0_{\mu_1 \ldots \mu_n} =
          {\cal O}^0_{\mu_{p(1)} \ldots \mu_{p(n)}}  \,, \, p \in S_n \,.
\end{equation}
Hence the operators are symmetric or antisymmetric in their indices,
or they transform according to some higher-dimensional irreducible
representation of $S_n$. Symmetrisation will be denoted by $\{ \cdots \}$,
i.e.
\begin{equation}
{\cal O}_{\{\mu_1 \mu_2 \ldots \mu_n \}} =
\frac{1}{n!} \sum_{p \in S_n}
{\cal O}_{\mu_{p(1)} \mu_{p(2)} \ldots \mu_{p(n)} } \,.
\end{equation}

\section{Irreducible Bases}

One finds the following bases transforming irreducibly under $\HG$.

\subsection{$n=1$}

$\tau^{(4)}_1$, $C=-1$:
\begin{equation}
{\cal O}^0_{\mu}\,, 1 \leq \mu \leq 4 \,,
\end{equation}

\subsection{$n=2$}

$\mbox{trace} \not = 0$, symmetric, $\tau^{(1)}_1$, $C=+1$:
\begin{equation}
\half \sum^4_{\mu = 1} {\cal O}^0_{\mu \mu}\,,
\end{equation}
$\mbox{trace} = 0$, symmetric, $\tau^{(3)}_1$, $C=+1$:
\begin{equation} \begin{array}{c} \displaystyle
\half \left({\cal O}^0_{1 1} + {\cal O}^0_{2 2} -
{\cal O}^0_{3 3} - {\cal O}^0_{4 4}    \right) \,,
\\ \displaystyle
\frac{1}{\sqrt{2}} \left({\cal O}^0_{3 3} - {\cal O}^0_{4 4}
   \right) \,,
\\ \displaystyle
\frac{1}{\sqrt{2}} \left({\cal O}^0_{1 1} - {\cal O}^0_{2 2}
   \right) \,,         \end{array}
\end{equation}
$\mbox{trace} = 0$, symmetric, $\tau^{(6)}_3$, $C=+1$:
\begin{equation}
\frac{1}{\sqrt{2}} \left({\cal O}^0_{\mu \nu} + {\cal O}^0_{\nu \mu}
   \right) \,, 1 \leq \mu < \nu \leq 4 \,,
\end{equation}
$\mbox{trace} = 0$, antisymmetric, $\tau^{(6)}_1$, $C=+1$:
\begin{equation}
\frac{1}{\sqrt{2}} \left({\cal O}^0_{\mu \nu} - {\cal O}^0_{\nu \mu}
   \right) \,, 1 \leq \mu < \nu \leq 4 \,.
\end{equation}

\subsection{$n=3$}

In order to simplify the notation we define for the case $n=3$
\begin{equation}
 {\cal O}_{|\nu_1 \nu_2 \nu_3 |} =
{\cal O}_{\nu_1 \nu_2 \nu_3 } - {\cal O}_{\nu_1 \nu_3 \nu_2 } -
{\cal O}_{\nu_3 \nu_1 \nu_2 } + {\cal O}_{\nu_3 \nu_2 \nu_1 } \,,
\end{equation}
\begin{equation}
 {\cal O}_{||\nu_1 \nu_2 \nu_3||} =
{\cal O}_{\nu_1 \nu_2 \nu_3 } - {\cal O}_{\nu_1 \nu_3 \nu_2 } +
{\cal O}_{\nu_3 \nu_1 \nu_2 } - {\cal O}_{\nu_3 \nu_2 \nu_1 } -
2 {\cal O}_{\nu_2 \nu_3 \nu_1 } + 2 {\cal O}_{\nu_2 \nu_1 \nu_3 } \,,
\end{equation}
\begin{equation}
 {\cal O}_{\langle \nu_1 \nu_2 \nu_3 \rangle} =
{\cal O}_{\nu_1 \nu_2 \nu_3 } + {\cal O}_{\nu_1 \nu_3 \nu_2 } +
{\cal O}_{\nu_3 \nu_1 \nu_2 } + {\cal O}_{\nu_3 \nu_2 \nu_1 } -
2 {\cal O}_{\nu_2 \nu_3 \nu_1 } - 2 {\cal O}_{\nu_2 \nu_1 \nu_3 } \,,
\end{equation}
\begin{equation}
 {\cal O}_{\langle \langle \nu_1 \nu_2 \nu_3 \rangle \rangle} =
{\cal O}_{\nu_1 \nu_2 \nu_3 } + {\cal O}_{\nu_1 \nu_3 \nu_2 } -
{\cal O}_{\nu_3 \nu_1 \nu_2 } - {\cal O}_{\nu_3 \nu_2 \nu_1 }
\end{equation}
and get the following bases transforming irreducibly under $\HG$.

\noindent
$\mbox{trace} \not = 0$, symmetric, $\tau^{(4)}_1$, $C=-1$:
\begin{equation}
\frac{1}{\sqrt{2}} \sum^4_{\nu = 1}
  {\cal O}^0_{\{\mu \nu \nu\}} \,, 1 \leq \mu \leq 4 \,,
\end{equation}
$\mbox{trace} = 0$, symmetric, $\tau^{(4)}_1$, $C=-1$:
\begin{equation}
\renewcommand{\arraystretch}{0.5}
\frac{1}{\sqrt{2}} \left (  {\cal O}^0_{\{\mu \mu \mu\}} -
          \sum^4_{\begin{array}{c}  \scriptstyle \nu = 1
                             \\ \scriptstyle \nu \not = \mu
                                             \end{array} }
  {\cal O}^0_{\{\mu \nu \nu\}} \right )
                            \,, 1 \leq \mu \leq 4 \,,
\end{equation}
$\mbox{trace} = 0$, symmetric, $\tau^{(8)}_1$, $C=-1$:
\begin{equation}
\begin{array}{ll} \displaystyle
\frac{\sqrt{3}}{\sqrt{2}} \left(
            {\cal O}^0_{\{122\}} - {\cal O}^0_{\{133\}} \right)  \,, &
\displaystyle
\frac{1}{\sqrt{2}} \left(
            {\cal O}^0_{\{122\}} + {\cal O}^0_{\{133\}} -
                           2 {\cal O}^0_{\{144\}} \right) \,,
                                             \\  \displaystyle
\frac{\sqrt{3}}{\sqrt{2}} \left(
            {\cal O}^0_{\{211\}} - {\cal O}^0_{\{233\}} \right)  \,, &
\displaystyle
\frac{1}{\sqrt{2}} \left(
            {\cal O}^0_{\{211\}} + {\cal O}^0_{\{233\}} -
                           2 {\cal O}^0_{\{244\}} \right) \,,
                                             \\  \displaystyle
\frac{\sqrt{3}}{\sqrt{2}} \left(
            {\cal O}^0_{\{311\}} - {\cal O}^0_{\{322\}} \right)  \,, &
\displaystyle
\frac{1}{\sqrt{2}} \left(
            {\cal O}^0_{\{311\}} + {\cal O}^0_{\{322\}} -
                           2 {\cal O}^0_{\{344\}} \right) \,,
                                             \\  \displaystyle
\frac{\sqrt{3}}{\sqrt{2}} \left(
            {\cal O}^0_{\{411\}} - {\cal O}^0_{\{422\}} \right)  \,, &
\displaystyle
\frac{1}{\sqrt{2}} \left(
            {\cal O}^0_{\{411\}} + {\cal O}^0_{\{422\}} -
                           2 {\cal O}^0_{\{433\}} \right) \,,
\end{array}
\end{equation}
$\mbox{trace} = 0$, symmetric, $\tau^{(4)}_2$, $C=-1$:
\begin{equation}
\sqrt{6} {\cal O}^0_{\{234\}} \,, \sqrt{6} {\cal O}^0_{\{134\}} \,,
\sqrt{6} {\cal O}^0_{\{124\}} \,, \sqrt{6} {\cal O}^0_{\{123\}} \,,
\end{equation}
$\mbox{trace} = 0$, antisymmetric, $\tau^{(4)}_4$, $C=+1$:
\begin{equation}
\frac{1}{\sqrt{6}} \sum_{p \in S_3} \mbox{sgn}(p)
{\cal O}^0_{\nu_{p(1)} \nu_{p(2)} \nu_{p(3)} }
\,, \nu_1 < \nu_2 < \nu_3 \,,
\end{equation}
$\mbox{trace} = 0$, mixed symmetry, $\tau^{(8)}_2$, $C=-1$:
\begin{equation}
\half {\cal O}^0_{\langle \langle \nu_1 \nu_2 \nu_3 \rangle \rangle
                                                             } \,,
 \frac{1}{2 \sqrt{3}}
      {\cal O}^0_{\langle \nu_1 \nu_2 \nu_3 \rangle} \,,
  \nu_1 < \nu_2 < \nu_3 \,,
\end{equation}
$\mbox{trace} = 0$, mixed symmetry, $\tau^{(8)}_2$, $C=+1$:
\begin{equation}
 \frac{1}{2 \sqrt{3}}
      {\cal O}^0_{|| \nu_1 \nu_2 \nu_3 ||} \,,
-\half {\cal O}^0_{| \nu_1 \nu_2 \nu_3 |} \,,
   \nu_1 < \nu_2 < \nu_3 \,,
\end{equation}
$\mbox{trace} \not = 0$, mixed symmetry, $\tau^{(4)}_1$, $C=-1$:
\begin{equation}
\renewcommand{\arraystretch}{0.5}
\frac{1}{3 \sqrt{2}}
 \sum^4_{\begin{array}{c}  \scriptstyle \nu = 1
                    \\ \scriptstyle \nu \not = \mu
                                       \end{array} }
{\cal O}^0_{\langle \langle \mu \nu \nu \rangle \rangle}
                            \,, 1 \leq \mu \leq 4 \,,
\end{equation}
$\mbox{trace} \not = 0$, mixed symmetry, $\tau^{(4)}_1$, $C=+1$:
\begin{equation}
\renewcommand{\arraystretch}{0.5}
\frac{1}{3 \sqrt{6}}
 \sum^4_{\begin{array}{c}  \scriptstyle \nu = 1
                    \\ \scriptstyle \nu \not = \mu
                                       \end{array} }
{\cal O}^0_{|| \mu \nu \nu ||} \,, 1 \leq \mu \leq 4 \,,
\end{equation}
$\mbox{trace} = 0$, mixed symmetry, $\tau^{(8)}_1$, $C=-1$:
\begin{equation}
\begin{array}{ll} \displaystyle
\frac{1}{2 \sqrt{3}} \left(
      {\cal O}^0_{\langle \langle 122 \rangle \rangle}
     -{\cal O}^0_{\langle \langle 133 \rangle \rangle} \right)  \,, &
\displaystyle
\frac{1}{6} \left(
      {\cal O}^0_{\langle \langle 122 \rangle \rangle}
     +{\cal O}^0_{\langle \langle 133 \rangle \rangle}
    -2{\cal O}^0_{\langle \langle 144 \rangle \rangle}  \right)  \,,
                                             \\  \displaystyle
\frac{1}{2 \sqrt{3}} \left(
      {\cal O}^0_{\langle \langle 211 \rangle \rangle}
     -{\cal O}^0_{\langle \langle 233 \rangle \rangle} \right)  \,, &
\displaystyle
\frac{1}{6} \left(
      {\cal O}^0_{\langle \langle 211 \rangle \rangle}
     +{\cal O}^0_{\langle \langle 233 \rangle \rangle}
    -2{\cal O}^0_{\langle \langle 244 \rangle \rangle}  \right)  \,,
                                             \\  \displaystyle
\frac{1}{2 \sqrt{3}} \left(
      {\cal O}^0_{\langle \langle 311 \rangle \rangle}
     -{\cal O}^0_{\langle \langle 322 \rangle \rangle} \right)  \,, &
\displaystyle
\frac{1}{6} \left(
      {\cal O}^0_{\langle \langle 311 \rangle \rangle}
     +{\cal O}^0_{\langle \langle 322 \rangle \rangle}
    -2{\cal O}^0_{\langle \langle 344 \rangle \rangle}  \right)  \,,
                                             \\  \displaystyle
\frac{1}{2 \sqrt{3}} \left(
      {\cal O}^0_{\langle \langle 411 \rangle \rangle}
     -{\cal O}^0_{\langle \langle 422 \rangle \rangle} \right)  \,, &
\displaystyle
\frac{1}{6} \left(
      {\cal O}^0_{\langle \langle 411 \rangle \rangle}
     +{\cal O}^0_{\langle \langle 422 \rangle \rangle}
    -2{\cal O}^0_{\langle \langle 433 \rangle \rangle}  \right)  \,,
\end{array}
\end{equation}
$\mbox{trace} = 0$, mixed symmetry, $\tau^{(8)}_1$, $C=+1$:
\begin{equation}
\label{mix0}
\begin{array}{ll} \displaystyle
\frac{1}{6} \left(
      {\cal O}^0_{|| 122 ||} - {\cal O}^0_{|| 133 ||} \right)  \,, &
\displaystyle
\frac{1}{6 \sqrt{3}} \left(
      {\cal O}^0_{|| 122 ||} + {\cal O}^0_{|| 133 ||}
    -2{\cal O}^0_{|| 144 ||}  \right)  \,,
                                             \\  \displaystyle
\frac{1}{6} \left(
      {\cal O}^0_{|| 211 ||} - {\cal O}^0_{|| 233 ||} \right)  \,, &
\displaystyle
\frac{1}{6 \sqrt{3}} \left(
      {\cal O}^0_{|| 211 ||} + {\cal O}^0_{|| 233 ||}
    -2{\cal O}^0_{|| 244 ||}  \right)  \,,
                                             \\  \displaystyle
\frac{1}{6} \left(
      {\cal O}^0_{|| 311 ||} - {\cal O}^0_{|| 322 ||} \right)  \,, &
\displaystyle
\frac{1}{6 \sqrt{3}} \left(
      {\cal O}^0_{|| 311 ||} + {\cal O}^0_{|| 322 ||}
    -2{\cal O}^0_{|| 344 ||}  \right)  \,,
                                             \\  \displaystyle
\frac{1}{6} \left(
      {\cal O}^0_{|| 411 ||} - {\cal O}^0_{|| 422 ||} \right)  \,, &
\displaystyle
\frac{1}{6 \sqrt{3}} \left(
      {\cal O}^0_{|| 411 ||} + {\cal O}^0_{|| 422 ||}
    -2{\cal O}^0_{|| 433 ||}  \right)  \,,
\end{array}
\end{equation}
Note that the operators with mixed symmetry and $C=+1$ can be
obtained from those with mixed symmetry and $C=-1$ by applying
\begin{equation}
\frac{1}{\sqrt{3}} \Bigl( P(\mbox{id}) + 2 P((12)) \Bigr) \,.
\end{equation}

\subsection{$n=4$}

For the totally symmetric tensors of rank 4 one can find bases for
irreducible subspaces by inspection.
For the not totally symmetric
rank-4 tensors we first make use of the embedding
$\HG \subset O(4) \subset GL(4)$ and start by decomposing with
respect to $GL(4)$. This can be done by standard textbook methods
(see, e.g., \cite{miller}). For $S_4$ we have five Young frames:
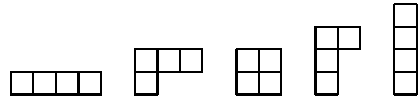
\begin{displaymath}
\symm \hspace{0.3cm} \liha \hspace{0.3cm} \kast \hspace{0.3cm}
\stha \hspace{0.3cm} \anti
\end{displaymath}
with the first one characterizing the totally symmetric tensors.
The corresponding irreducible representations of $GL(4)$ have the
dimensions
\begin{displaymath}
35 \qquad 45 \qquad 20 \qquad 15 \qquad 1
\end{displaymath}
and occur with multiplicity
\begin{displaymath}
1  \qquad  3 \qquad 2  \qquad 3  \qquad 1 \,.
\end{displaymath}
The subspaces of traceless tensors, which carry the irreducible
representation of $O(4)$ corresponding to the particular Young
frame, have the dimensions
\begin{displaymath}
25 \qquad 30 \qquad 10 \qquad 9  \qquad 1 \,.
\end{displaymath}

For each of these irreducible representations of $GL(4)$, except for
that on the totally symmetric tensors, we construct, following the
standard recipes, a projector on an irreducible subspace consisting
of tensors which are antisymmetric upon interchange of the second
and the fourth index. (Interchange of the second and the fourth index
corresponds to charge conjugation, cf.\ (\ref{chacon}).)
The further decomposition into $\HG$-irreducible subspaces is
then performed by inspection. In this process the knowledge of the
above mentioned dimensions proves rather valuable.

In the case of the $GL(4)$ representations which occur with
multiplicity greater than one we still need bases for the other
irreducible subspaces carrying equivalent $GL(4)$ representations.
We get them by acting on the basis
already constructed with suitable operators of the form
\begin{equation}
\sum_{p \in S_4} a(p) P(p)
\end{equation}
where $P(p)$ was defined in (\ref{symmact}).
The basis obtained in this way has the same transformation properties
with respect to $GL(4)$ (and hence also with respect to $\HG$) as the
basis we started with, because the actions of the symmetric group and
of the hypercubic group commute. Furthermore the coefficients $a(p)$
will be chosen such that the basis elements are eigenvectors
of $(24) \in S_4$.

\subsubsection[xcv]{\symm}

We begin by listing $\HG$-irreducible bases in the space of totally
symmetric tensors belonging to the Young frame \symm.

\noindent
$\mbox{trace} \not = 0$, $\tau^{(1)}_1$, $C=+1$:
\begin{equation}
\renewcommand{\arraystretch}{0.5}
\frac{1}{2 \sqrt{2}}  \sum^4_{\mu = 1}
                     {\cal O}^0_{\{\mu \mu \mu \mu \}} +
\frac{1}{\sqrt{2}}
 \sum^4_{\begin{array}{c}  \scriptstyle \mu , \nu = 1
                    \\ \scriptstyle \mu < \nu
                                       \end{array} }
                     {\cal O}^0_{\{\mu \mu \nu \nu \}} \,,
\end{equation}
$\mbox{trace} \not = 0$, $\tau^{(3)}_1$, $C=+1$:
\begin{equation} \begin{array}{c} \displaystyle
\frac{\sqrt{3}}{4} \left({\cal O}^0_{\{1111\}} + {\cal O}^0_{\{2222\}} -
{\cal O}^0_{\{3333\}} - {\cal O}^0_{\{4444\}}    \right)    +
\frac{\sqrt{3}}{2} \left({\cal O}^0_{\{1122\}} - {\cal O}^0_{\{3344\}}
                                                 \right) \,,
\\ \displaystyle
\frac{\sqrt{3}}{2 \sqrt{2}} \left(
{\cal O}^0_{\{3333\}} - {\cal O}^0_{\{4444\}}    \right)    +
\frac{\sqrt{3}}{2 \sqrt{2}} \left(
{\cal O}^0_{\{1133\}} - {\cal O}^0_{\{2244\}} -
{\cal O}^0_{\{1144\}} + {\cal O}^0_{\{2233\}}
                                                 \right) \,,
\\ \displaystyle
\frac{\sqrt{3}}{2 \sqrt{2}} \left(
{\cal O}^0_{\{1111\}} - {\cal O}^0_{\{2222\}}    \right)    +
\frac{\sqrt{3}}{2 \sqrt{2}} \left(
{\cal O}^0_{\{1133\}} - {\cal O}^0_{\{2244\}} +
{\cal O}^0_{\{1144\}} - {\cal O}^0_{\{2233\}}
                                                 \right) \,,
\end{array}
\end{equation}
$\mbox{trace} \not = 0$, $\tau^{(6)}_3$, $C=+1$:
\begin{equation}
\frac{\sqrt{3}}{\sqrt{2}}  \left(
 {\cal O}^0_{\{\mu \mu \mu \nu \}} +
 {\cal O}^0_{\{\nu \nu \nu \mu \}} +
 {\cal O}^0_{\{\nu_1 \nu_1 \mu \nu \}} +
 {\cal O}^0_{\{\nu_2 \nu_2 \mu \nu \}}  \right)
 \,, 1 \leq \mu < \nu \leq 4 \,,
\end{equation}
where $\nu_1 \,, \nu_2$ are such that $\nu_1 < \nu_2$ and
$\{\mu,\nu,\nu_1,\nu_2\} = \{1,2,3,4\}$,

\noindent
$\mbox{trace} = 0$, $\tau^{(1)}_1$, $C=+1$:
\begin{equation}
\renewcommand{\arraystretch}{0.5}
\frac{1}{2 \sqrt{2}}  \sum^4_{\mu = 1}
                     {\cal O}^0_{\{\mu \mu \mu \mu \}} -
\frac{1}{\sqrt{2}}
 \sum^4_{\begin{array}{c}  \scriptstyle \mu ,\nu = 1
                    \\ \scriptstyle \mu < \nu
                                       \end{array} }
                     {\cal O}^0_{\{\mu \mu \nu \nu \}} \,,
\end{equation}
$\mbox{trace} = 0$, $\tau^{(3)}_1$, $C=+1$:
\begin{equation} \begin{array}{c} \displaystyle
\frac{1}{4} \left({\cal O}^0_{\{1111\}} + {\cal O}^0_{\{2222\}} -
{\cal O}^0_{\{3333\}} - {\cal O}^0_{\{4444\}}    \right)    -
\frac{3}{2} \left({\cal O}^0_{\{1122\}} - {\cal O}^0_{\{3344\}}
                                                 \right) \,,
\\ \displaystyle
\frac{1}{2 \sqrt{2}} \left(
{\cal O}^0_{\{3333\}} - {\cal O}^0_{\{4444\}}    \right)    -
\frac{3 }{2 \sqrt{2}} \left(
{\cal O}^0_{\{1133\}} - {\cal O}^0_{\{2244\}} -
{\cal O}^0_{\{1144\}} + {\cal O}^0_{\{2233\}}
                                                 \right) \,,
\\ \displaystyle
\frac{1}{2 \sqrt{2}} \left(
{\cal O}^0_{\{1111\}} - {\cal O}^0_{\{2222\}}    \right)    -
\frac{3}{2 \sqrt{2}} \left(
{\cal O}^0_{\{1133\}} - {\cal O}^0_{\{2244\}} +
{\cal O}^0_{\{1144\}} - {\cal O}^0_{\{2233\}}
                                                 \right) \,,
\end{array}
\end{equation}
$\mbox{trace} = 0$, $\tau^{(6)}_3$, $C=+1$:
\begin{equation}
\frac{1}{\sqrt{2}}  \left(
 {\cal O}^0_{\{\mu \mu \mu \nu \}} +
 {\cal O}^0_{\{\nu \nu \nu \mu \}} \right) -
\frac{3}{\sqrt{2}}  \left(
 {\cal O}^0_{\{\nu_1 \nu_1 \mu \nu \}} +
 {\cal O}^0_{\{\nu_2 \nu_2 \mu \nu \}}  \right)
 \,, 1 \leq \mu < \nu \leq 4 \,,
\end{equation}
where $\nu_1 \,, \nu_2$ are such that $\nu_1 < \nu_2$ and
$\{\mu,\nu,\nu_1,\nu_2\} = \{1,2,3,4\}$,

\noindent
$\mbox{trace} = 0$, $\tau^{(2)}_1$, $C=+1$:
\begin{equation} \begin{array}{c} \displaystyle
\frac{\sqrt{3}}{\sqrt{2}} \left(
{\cal O}^0_{\{1122\}} + {\cal O}^0_{\{3344\}} -
{\cal O}^0_{\{1133\}} - {\cal O}^0_{\{2244\}}    \right) \,,
\\ \displaystyle
- \frac{1}{\sqrt{2}} \left(
{\cal O}^0_{\{1122\}} + {\cal O}^0_{\{3344\}} +
{\cal O}^0_{\{1133\}} + {\cal O}^0_{\{2244\}} -
2 {\cal O}^0_{\{1144\}} - 2 {\cal O}^0_{\{2233\}}    \right) \,,
\end{array}
\end{equation}
$\mbox{trace} = 0$, $\tau^{(1)}_2$, $C=+1$:
\begin{equation}
2 \sqrt{6}  {\cal O}^0_{\{1234\}}  \,,
\end{equation}
$\mbox{trace} = 0$, $\tau^{(6)}_1$, $C=+1$:
\begin{equation}
 \sqrt{2}  \left( {\cal O}^0_{\{\mu \mu \mu \nu \}} -
 {\cal O}^0_{\{\nu \nu \nu \mu \}} \right)
 \,, 1 \leq \mu < \nu \leq 4 \,,
\end{equation}
$\mbox{trace} = 0$, $\tau^{(6)}_2$, $C=+1$:
\begin{equation}
 \sqrt{6}  \left( {\cal O}^0_{\{\nu_1 \nu_1 \mu \nu \}} -
 {\cal O}^0_{\{\nu_2 \nu_2 \mu \nu \}}  \right)
 \,, 1 \leq \mu < \nu \leq 4 \,,
\end{equation}
where $\nu_1 \,, \nu_2$ are such that $\nu_1 < \nu_2$ and
$\{\mu,\nu,\nu_1,\nu_2\} = \{1,2,3,4\}$.

\subsubsection[xcv]{\raisebox{-0.15cm}{\liha}}

In order to describe a basis for an irreducible subspace
(with respect to $GL(4)$) corresponding to \liha we define
\begin{equation}
 {\cal O}^{\mu \nu}_{\lambda \rho} = N_{\lambda \rho}
 \left ( {\cal O}^0_{\lambda \mu \rho \nu}
       - {\cal O}^0_{\lambda \nu \rho \mu}
       + {\cal O}^0_{\rho \mu \lambda \nu}
       - {\cal O}^0_{\rho \nu \lambda \mu} \right )
\end{equation}
with
\begin{equation}
\renewcommand{\arraystretch}{0.5}
 N_{\lambda \rho} = \left \{
 \begin{array}{lcl}
      1/2         & \mbox{for} & \lambda \neq \rho \,,  \\
      1/\sqrt{8}  & \mbox{for} & \lambda = \rho \,.
 \end{array} \right.
\end{equation}
With respect to $\HG$ one gets the following irreducible subspaces
of this $GL(4)$-irreducible space.

\noindent
$\mbox{trace} \not = 0$, $\tau^{(3)}_1$, $C=-1$:
\begin{equation} \begin{array}{c} \displaystyle
\label{liha0}
\frac{1}{\sqrt{3}} \left(
{\cal O}^{12}_{12} - {\cal O}^{23}_{23} -
{\cal O}^{24}_{24}  \right) \,,
\\ \displaystyle
\frac{1}{2 \sqrt{6}} \left(
- {\cal O}^{12}_{12} - 3 {\cal O}^{13}_{13} - 2{\cal O}^{23}_{23}
+ {\cal O}^{24}_{24} + 3 {\cal O}^{34}_{34} \right) \,,
\\ \displaystyle
\frac{1}{2 \sqrt{2}} \left(
  {\cal O}^{12}_{12} +   {\cal O}^{13}_{13} + 2 {\cal O}^{14}_{14}
+ {\cal O}^{24}_{24} +   {\cal O}^{34}_{34} \right) \,,
\end{array}
\end{equation}
$\mbox{trace} \not = 0$, $\tau^{(6)}_1$, $C=-1$:
\begin{equation}
- \frac{1}{\sqrt{3}} \left( {\cal O}^{\nu \lambda}_{\nu \nu}
        + {\cal O}^{\nu \lambda}_{\lambda \lambda} \right)
+ \frac{1}{2 \sqrt{6}} \sum^2_{i=1} \left(
               {\cal O}^{\mu_i \nu}_{\mu_i \lambda}
            -  {\cal O}^{\mu_i \lambda}_{\mu_i \nu}
     - \sqrt{2} {\cal O}^{\nu \lambda}_{\mu_i \mu_i} \right)
                    \,, \nu < \lambda \,,
\end{equation}
$\mbox{trace} \not = 0$, $\tau^{(6)}_3$, $C=-1$:
\begin{equation}
  \frac{1}{2} \left( {\cal O}^{\nu \lambda}_{\nu \nu}
        - {\cal O}^{\nu \lambda}_{\lambda \lambda} \right)
+ \frac{1}{2 \sqrt{2}} \sum^2_{i=1} \left(
               {\cal O}^{\mu_i \nu}_{\mu_i \lambda}
            +  {\cal O}^{\mu_i \lambda}_{\mu_i \nu}  \right)
                    \,, \nu < \lambda \,,
\end{equation}
$\mbox{trace} = 0$, $\tau^{(3)}_2$, $C=-1$:
\begin{equation} \begin{array}{c} \displaystyle
\frac{1}{\sqrt{3}} \left(
{\cal O}^{14}_{23} + {\cal O}^{24}_{13} +
{\cal O}^{34}_{12}  \right) \,,
\\ \displaystyle
\frac{1}{2 \sqrt{6}} \left(
3 {\cal O}^{13}_{24} +   {\cal O}^{14}_{23} + 3 {\cal O}^{23}_{14}
+ {\cal O}^{24}_{13} - 2 {\cal O}^{34}_{12} \right) \,,
\\ \displaystyle
\frac{1}{2 \sqrt{2}} \left(
2 {\cal O}^{12}_{34} +   {\cal O}^{13}_{24} +   {\cal O}^{14}_{23}
- {\cal O}^{23}_{14} -   {\cal O}^{24}_{13} \right) \,,
\end{array}
\end{equation}
$\mbox{trace} = 0$, $\tau^{(3)}_3$, $C=-1$:
\begin{equation} \begin{array}{c} \displaystyle
\frac{1}{\sqrt{3}} \left(
{\cal O}^{12}_{12} - {\cal O}^{13}_{13} +
{\cal O}^{23}_{23}  \right) \,,
\\ \displaystyle
\frac{1}{2 \sqrt{6}} \left(
-2 {\cal O}^{12}_{12} -   {\cal O}^{13}_{13} + 3 {\cal O}^{14}_{14}
+ {\cal O}^{23}_{23} - 3 {\cal O}^{24}_{24} \right) \,,
\\ \displaystyle
\frac{1}{2 \sqrt{2}} \left(
- {\cal O}^{13}_{13} +   {\cal O}^{14}_{14} - {\cal O}^{23}_{23}
+ {\cal O}^{24}_{24} - 2 {\cal O}^{34}_{34} \right) \,,
\end{array}
\end{equation}
$\mbox{trace} = 0$, $\tau^{(6)}_1$, $C=-1$:
\begin{equation}
  \frac{1}{\sqrt{6}} \left( {\cal O}^{\nu \lambda}_{\nu \nu}
        + {\cal O}^{\nu \lambda}_{\lambda \lambda} \right)
+ \frac{1}{2 \sqrt{3}} \sum^2_{i=1} \left(
               {\cal O}^{\mu_i \nu}_{\mu_i \lambda}
            -  {\cal O}^{\mu_i \lambda}_{\mu_i \nu}
     - \sqrt{2} {\cal O}^{\nu \lambda}_{\mu_i \mu_i} \right)
                    \,, \nu < \lambda \,,
\end{equation}
$\mbox{trace} = 0$, $\tau^{(6)}_2$, $C=-1$:
\begin{equation}
  \frac{1}{2} \sum^2_{i=1} (-1)^{i+1} \left(
               {\cal O}^{\mu_i \nu}_{\mu_i \lambda}
            +  {\cal O}^{\mu_i \lambda}_{\mu_i \nu}  \right)
                    \,, \nu < \lambda \,,
\end{equation}
$\mbox{trace} = 0$, $\tau^{(6)}_3$, $C=-1$:
\begin{equation}
  \frac{1}{2} \left( {\cal O}^{\nu \lambda}_{\nu \nu}
        - {\cal O}^{\nu \lambda}_{\lambda \lambda} \right)
- \frac{1}{2 \sqrt{2}} \sum^2_{i=1} \left(
               {\cal O}^{\mu_i \nu}_{\mu_i \lambda}
            +  {\cal O}^{\mu_i \lambda}_{\mu_i \nu}  \right)
                    \,, \nu < \lambda \,,
\end{equation}
$\mbox{trace} = 0$, $\tau^{(6)}_4$, $C=-1$:
\begin{equation}  \label{liha1}
  \frac{1}{2} \sum^2_{i=1} (-1)^{i+1} \left(
               {\cal O}^{\mu_i \nu}_{\mu_i \lambda}
            -  {\cal O}^{\mu_i \lambda}_{\mu_i \nu}
     - \sqrt{2} {\cal O}^{\nu \lambda}_{\mu_i \mu_i} \right)
                    \,, \nu < \lambda \,.
\end{equation}
For $\nu < \lambda$ the indices $\mu_1,\mu_2$ are such that
$\mu_1 < \mu_2$ and
$\{\nu,\lambda,\mu_1,\mu_2\} = \{1,2,3,4\}$.

Applying the operators
\begin{equation}
\frac{1}{\sqrt{3}} \Bigl( P(\mbox{id}) - 2 P((12)) \Bigr) \,,
\end{equation}
\begin{equation}
\frac{1}{\sqrt{6}} \Bigl( P(\mbox{id}) + P((12)) - 3 P((23))
                                                    \Bigr) \,,
\end{equation}
respectively, to the basis vectors (\ref{liha0}) - (\ref{liha1})
one obtains
an orthonormal system of vectors which transform under $\HG$
exactly like (\ref{liha0}) - (\ref{liha1}) but have $C=+1$.

\subsubsection[xcv]{\raisebox{-0.15cm}{\kast}}

In order to describe a basis for an irreducible subspace
(with respect to $GL(4)$) corresponding to this Young frame
we define
\begin{equation} \begin{array}{l} \displaystyle
 \tilde{{\cal O}}_{ij} = N_{ij}
 \left ( {\cal O}^0_{\mu_i \mu_j \nu_i \nu_j}
       - {\cal O}^0_{\nu_i \mu_j \mu_i \nu_j}
       - {\cal O}^0_{\mu_i \nu_j \nu_i \mu_j}
       + {\cal O}^0_{\nu_i \nu_j \mu_i \mu_j} \right.
 \\ \displaystyle
 \hphantom{ {\cal O}_{ij} = N_{ij} (}   \left.
    { }+ {\cal O}^0_{\mu_j \mu_i \nu_j \nu_i}
       - {\cal O}^0_{\nu_j \mu_i \mu_j \nu_i}
       - {\cal O}^0_{\mu_j \nu_i \nu_j \mu_i}
       + {\cal O}^0_{\nu_j \nu_i \mu_j \mu_i} \right)
\end{array}
\end{equation}
with
\begin{equation}
\renewcommand{\arraystretch}{0.5}
 N_{ij} = \left \{
 \begin{array}{lcl}
      1/\sqrt{8}  & \mbox{for} & i \neq j \,,  \\
      1/4         & \mbox{for} & i = j \,.
 \end{array} \right.
\end{equation}
The indices $i,j \in \{1,2,3,4,5,6 \}$
enumerate pairs $\mu \, \nu$ with $\mu < \nu$
according to the scheme
\begin{center}
\renewcommand{\arraystretch}{0.7}
\begin{tabular}{ccccccc}
 $i$     & 1 & 2 & 3 & 4 & 5 & 6 \\
 $\mu_i$ & 1 & 1 & 1 & 2 & 2 & 3 \\
 $\nu_i$ & 2 & 3 & 4 & 3 & 4 & 4
\end{tabular}
\end{center}
Note that $\tilde{{\cal O}}_{ij} = \tilde{{\cal O}}_{ji} $.

With respect to $\HG$ one gets the following irreducible subspaces
of the $GL(4)$-irreducible space.

\noindent
$\mbox{trace} \not = 0$, $\tau^{(1)}_1$, $C=-1$:
\begin{equation} \label{kast0}
\frac{1}{\sqrt{6}} \left(
 \tilde{{\cal O}}_{11} + \cdots + \tilde{{\cal O}}_{66} \right) \,,
\end{equation}
$\mbox{trace} \not = 0$, $\tau^{(3)}_1$, $C=-1$:
\begin{equation}
\frac{1}{\sqrt{2}} \left(
 \tilde{{\cal O}}_{11} - \tilde{{\cal O}}_{66} \right) \,,
\frac{1}{\sqrt{2}} \left(
 \tilde{{\cal O}}_{22} - \tilde{{\cal O}}_{55} \right) \,,
\frac{1}{\sqrt{2}} \left(
 \tilde{{\cal O}}_{33} - \tilde{{\cal O}}_{44} \right) \,,
\end{equation}
$\mbox{trace} \not = 0$, $\tau^{(6)}_3$, $C=-1$:
\begin{equation} \begin{array}{cccccc} \displaystyle
\frac{1}{\sqrt{2}} \left(
 \tilde{{\cal O}}_{12} + \tilde{{\cal O}}_{56} \right) &,& \displaystyle
\frac{1}{\sqrt{2}} \left(
 \tilde{{\cal O}}_{13} - \tilde{{\cal O}}_{46} \right) &,& \displaystyle
\frac{1}{\sqrt{2}} \left(
 \tilde{{\cal O}}_{14} - \tilde{{\cal O}}_{36} \right) & ,
\\ \displaystyle
\frac{1}{\sqrt{2}} \left(
 \tilde{{\cal O}}_{15} + \tilde{{\cal O}}_{26} \right) &,& \displaystyle
\frac{1}{\sqrt{2}} \left(
 \tilde{{\cal O}}_{23} + \tilde{{\cal O}}_{45} \right) &,& \displaystyle
\frac{1}{\sqrt{2}} \left(
 \tilde{{\cal O}}_{24} + \tilde{{\cal O}}_{35} \right) & ,
\end{array}
\end{equation}
$\mbox{trace} = 0$, $\tau^{(2)}_1$, $C=-1$:
\begin{equation}
\frac{1}{\sqrt{12}} \left(
  -2 \tilde{{\cal O}}_{11} -2 \tilde{{\cal O}}_{66}
   + \tilde{{\cal O}}_{22} +  \tilde{{\cal O}}_{33}
   + \tilde{{\cal O}}_{44} +  \tilde{{\cal O}}_{55}     \right) \,,
\frac{1}{2} \left(
     \tilde{{\cal O}}_{33} +  \tilde{{\cal O}}_{44}
   - \tilde{{\cal O}}_{22} -  \tilde{{\cal O}}_{55}     \right) \,,
\end{equation}
$\mbox{trace} = 0$, $\tau^{(2)}_2$, $C=-1$:
\begin{equation}
\frac{1}{\sqrt{2}} \left(
     \tilde{{\cal O}}_{16} +  \tilde{{\cal O}}_{25}  \right) \,,
\frac{1}{\sqrt{6}} \left(
     \tilde{{\cal O}}_{16} -  \tilde{{\cal O}}_{25}
  -2 \tilde{{\cal O}}_{34} \right) \,,
\end{equation}
$\mbox{trace} = 0$, $\tau^{(6)}_2$, $C=-1$:
\begin{equation} \label{kast1} \begin{array}{cccccc} \displaystyle
\frac{1}{\sqrt{2}} \left(
     \tilde{{\cal O}}_{12} - \tilde{{\cal O}}_{56} \right)
                                               & , & \displaystyle
\frac{1}{\sqrt{2}} \left(
     \tilde{{\cal O}}_{13} + \tilde{{\cal O}}_{46} \right)
                                               & , & \displaystyle
\frac{1}{\sqrt{2}} \left(
     \tilde{{\cal O}}_{14} + \tilde{{\cal O}}_{36} \right) & ,
\\ \displaystyle
\frac{1}{\sqrt{2}} \left(
     \tilde{{\cal O}}_{15} - \tilde{{\cal O}}_{26} \right)
                                               & , & \displaystyle
\frac{1}{\sqrt{2}} \left(
     \tilde{{\cal O}}_{23} - \tilde{{\cal O}}_{45} \right)
                                               & , & \displaystyle
\frac{1}{\sqrt{2}} \left(
     \tilde{{\cal O}}_{24} - \tilde{{\cal O}}_{35} \right) & .
\end{array}
\end{equation}

Applying the operator
\begin{equation}
\frac{1}{\sqrt{3}} \Bigl( P(\mbox{id}) - 2 P((12)) \Bigr)
\end{equation}
to the basis vectors (\ref{kast0}) - (\ref{kast1})
one obtains
an orthonormal system of vectors which transform under $\HG$
exactly like (\ref{kast0}) - (\ref{kast1}) but have $C=+1$.

\subsubsection[xcv]{\raisebox{-0.3cm}{\stha}}

In order to describe a basis for an irreducible subspace
(with respect to $GL(4)$) corresponding to this Young frame
we define
\begin{equation}
 \hat{{\cal O}}_{\mu \nu} = \frac{1}{\sqrt{6}}
     \sum_{p \in S_3}  \mbox{sgn}(p)
         {\cal O}^0_{\nu_{p(1)} \nu_{p(2)} \mu \, \nu_{p(3)}} \,.
\end{equation}
The indices $\nu_1$, $\nu_2$, $\nu_3$ are determined by $\nu$ through
the conditions $\nu_1 < \nu_2 < \nu_3$ and
$\{ \nu_1,\nu_2,\nu_3,\nu \} = \{1,2,3,4\}$.

With respect to $\HG$ one gets the following irreducible subspaces
of the $GL(4)$-irreducible space.

\noindent
$\mbox{trace} \not = 0$, $\tau^{(6)}_1$, $C=-1$:
\begin{equation} \label{stha0} \begin{array}{cccccc} \displaystyle
\frac{1}{\sqrt{2}} \left(
     \hat{{\cal O}}_{12} + \hat{{\cal O}}_{21} \right)
                                               & , & \displaystyle
\frac{1}{\sqrt{2}} \left(
     \hat{{\cal O}}_{13} - \hat{{\cal O}}_{31} \right)
                                               & , & \displaystyle
\frac{1}{\sqrt{2}} \left(
     \hat{{\cal O}}_{14} + \hat{{\cal O}}_{41} \right) & ,
\\ \displaystyle
\frac{1}{\sqrt{2}} \left(
     \hat{{\cal O}}_{23} + \hat{{\cal O}}_{32} \right)
                                               & , & \displaystyle
\frac{1}{\sqrt{2}} \left(
     \hat{{\cal O}}_{24} - \hat{{\cal O}}_{42} \right)
                                               & , & \displaystyle
\frac{1}{\sqrt{2}} \left(
     \hat{{\cal O}}_{34} + \hat{{\cal O}}_{43} \right) & ,
\end{array}
\end{equation}
$\mbox{trace} = 0$, $\tau^{(3)}_4$, $C=-1$:
\begin{equation}
\frac{1}{2} \left(
     \hat{{\cal O}}_{11} - \hat{{\cal O}}_{22}
   - \hat{{\cal O}}_{33} + \hat{{\cal O}}_{44} \right) \,,
\frac{1}{\sqrt{2}} \left(
     \hat{{\cal O}}_{33} + \hat{{\cal O}}_{44} \right) \,,
\frac{1}{\sqrt{2}} \left(
     \hat{{\cal O}}_{11} + \hat{{\cal O}}_{22} \right) \,,
\end{equation}
$\mbox{trace} = 0$, $\tau^{(6)}_4$, $C=-1$:
\begin{equation} \label{stha1} \begin{array}{cccccc} \displaystyle
\frac{1}{\sqrt{2}} \left(
     \hat{{\cal O}}_{12} - \hat{{\cal O}}_{21} \right)
                                               & , & \displaystyle
\frac{1}{\sqrt{2}} \left(
     \hat{{\cal O}}_{13} + \hat{{\cal O}}_{31} \right)
                                               & , & \displaystyle
\frac{1}{\sqrt{2}} \left(
     \hat{{\cal O}}_{14} - \hat{{\cal O}}_{41} \right) & ,
\\ \displaystyle
\frac{1}{\sqrt{2}} \left(
     \hat{{\cal O}}_{23} - \hat{{\cal O}}_{32} \right)
                                               & , & \displaystyle
\frac{1}{\sqrt{2}} \left(
     \hat{{\cal O}}_{24} + \hat{{\cal O}}_{42} \right)
                                               & , & \displaystyle
\frac{1}{\sqrt{2}} \left(
     \hat{{\cal O}}_{34} - \hat{{\cal O}}_{43} \right) & .
\end{array}
\end{equation}

Applying the operators
\begin{equation} \label{opminus}
\frac{1}{2 \sqrt{2}} \Bigl( 2 P(\mbox{id}) - 3 P((23)) - 3 P((34))
                                                        \Bigr) \,,
\end{equation}
\begin{equation} \label{opplus}
\sqrt{\frac{3}{8}} \Bigl( P((23)) - P((34)) \Bigr) \,,
\end{equation}
respectively, to the basis vectors (\ref{stha0}) - (\ref{stha1})
one obtains two
orthonormal systems of vectors which transform under $\HG$
exactly like (\ref{stha0}) - (\ref{stha1}). Using (\ref{opminus})
one gets $C=-1$, whereas (\ref{opplus}) leads to $C=+1$.

\subsubsection[xcv]{\raisebox{-0.45cm}{\anti}}

The irreducible subspace corresponding to this Young frame is
one-dimensional and carries the following representation
of $\HG$.

\noindent
$\mbox{trace} = 0$, $\tau^{(1)}_4$, $C=-1$:
\begin{equation}
  \frac{1}{\sqrt{24}}      \sum_{p \in S_4}   \mbox{sgn}(p)
         {\cal O}^0_{p(1) \cdots p(4)} \,.
\end{equation}

\section{Axial Operators}

The transformation law (\ref{trans5}) of the operators
${\cal O}^5_{\mu_1 \mu_2 \ldots \mu_n}$
differs from the transformation law (\ref{trans0}) only by the
additional factor of $\det (T(a,\pi))$. Since the mapping
\begin{equation}
(a,\pi) \to  \det (T(a,\pi))
\end{equation}
is the (one-dimensional) irreducible representation $ \tau^{(1)}_4$
bases of irreducible subspaces can be constructed from the operators
${\cal O}^5_{\mu_1 \mu_2 \ldots \mu_n}$ in exactly the same manner
as from the operators ${\cal O}^0_{\mu_1 \mu_2 \ldots \mu_n}$.
The only difference is that $C$-parity changes sign and
the representations acting in these
subspaces have to be multiplied by $ \tau^{(1)}_4$. One finds
\begin{equation}
\renewcommand{\arraystretch}{1.4}
\begin{array}{ll} \displaystyle
 \tau^{(1)}_1  \to   \tau^{(1)}_4 \otimes  \tau^{(1)}_1
                           =  \tau^{(1)}_4 \,, &
 \tau^{(1)}_2  \to   \tau^{(1)}_4 \otimes  \tau^{(1)}_2
                           =  \tau^{(1)}_3 \,,
 \\ \displaystyle
 \tau^{(1)}_3  \to   \tau^{(1)}_4 \otimes  \tau^{(1)}_3
                           =  \tau^{(1)}_2 \,, &
 \tau^{(1)}_4  \to   \tau^{(1)}_4 \otimes  \tau^{(1)}_4
                           =  \tau^{(1)}_1 \,,
 \\ \displaystyle
 \tau^{(2)}_1  \to   \tau^{(1)}_4 \otimes  \tau^{(2)}_1
                           =  \tau^{(2)}_2 \,, &
 \tau^{(2)}_2  \to   \tau^{(1)}_4 \otimes  \tau^{(2)}_2
                           =  \tau^{(2)}_1 \,,
 \\ \displaystyle
 \tau^{(3)}_1  \to   \tau^{(1)}_4 \otimes  \tau^{(3)}_1
                           =  \tau^{(3)}_4 \,, &
 \tau^{(3)}_2  \to   \tau^{(1)}_4 \otimes  \tau^{(3)}_2
                           =  \tau^{(3)}_3 \,,
 \\ \displaystyle
 \tau^{(3)}_3  \to   \tau^{(1)}_4 \otimes  \tau^{(3)}_3
                           =  \tau^{(3)}_2 \,, &
 \tau^{(3)}_4  \to   \tau^{(1)}_4 \otimes  \tau^{(3)}_4
                           =  \tau^{(3)}_1 \,,
 \\ \displaystyle
 \tau^{(4)}_1  \to   \tau^{(1)}_4 \otimes  \tau^{(4)}_1
                           =  \tau^{(4)}_4 \,, &
 \tau^{(4)}_2  \to   \tau^{(1)}_4 \otimes  \tau^{(4)}_2
                           =  \tau^{(4)}_3 \,,
 \\ \displaystyle
 \tau^{(4)}_3  \to   \tau^{(1)}_4 \otimes  \tau^{(4)}_3
                           =  \tau^{(4)}_2 \,, &
 \tau^{(4)}_4  \to   \tau^{(1)}_4 \otimes  \tau^{(4)}_4
                           =  \tau^{(4)}_1 \,,
 \\ \displaystyle
 \tau^{(6)}_1  \to   \tau^{(1)}_4 \otimes  \tau^{(6)}_1
                           =  \tau^{(6)}_1 \,, &
 \tau^{(6)}_2  \to   \tau^{(1)}_4 \otimes  \tau^{(6)}_2
                           =  \tau^{(6)}_2 \,,
 \\ \displaystyle
 \tau^{(6)}_3  \to   \tau^{(1)}_4 \otimes  \tau^{(6)}_3
                           =  \tau^{(6)}_4 \,, &
 \tau^{(6)}_4  \to   \tau^{(1)}_4 \otimes  \tau^{(6)}_4
                           =  \tau^{(6)}_3 \,,
 \\ \displaystyle
 \tau^{(8)}_1  \to   \tau^{(1)}_4 \otimes  \tau^{(8)}_1
                           =  \tau^{(8)}_2 \,, &
 \tau^{(8)}_2  \to   \tau^{(1)}_4 \otimes  \tau^{(8)}_2
                           =  \tau^{(8)}_1 \,.
\end{array} \end{equation}

\section{Mixing}

As one sees, quite a number of pairs
$\left( \tau^{(l)}_k , C \right)$ occur more than once leading
to potential mixing problems. In order to find out which operators
could mix one needs bases for the equivalent subspaces which
transform {\em identically} under $\HG$. The bases given above
do not always have this property, but constructing appropriate
basis transformations should present no difficulties in specific
cases of interest. In this connection it is useful to remember
that $\HG$ is generated by the group elements
(\ref{genalpha}) -- (\ref{gengamma}) so that the transformation
behaviour of a given basis has to be checked only for these
three elements. We shall quote only a few examples relevant for
our own numerical work \cite{letter}, where
we have studied (among others) the
following operators and representations.

\noindent
$\mbox{trace} = 0$, symmetric, $\tau^{(8)}_1$, $C=-1$:
\begin{equation} \label{mix1}
\begin{array}{l} \displaystyle
 {\cal O}^0_{\{411\}} - \half \left( {\cal O}^0_{\{422\}} +
                    {\cal O}^0_{\{433\}} \right) \\
 \displaystyle \quad
  = \frac{\sqrt{3}}{2 \sqrt{2}} \cdot \frac{\sqrt{3}}{\sqrt{2}}
    \left( {\cal O}^0_{\{411\}} - {\cal O}^0_{\{422\}} \right) +
    \frac{1}{2 \sqrt{2}} \cdot \frac{1}{\sqrt{2}} \left(
     {\cal O}^0_{\{411\}} + {\cal O}^0_{\{422\}} -
   2 {\cal O}^0_{\{433\}} \right) \,.
\end{array}
\end{equation}
Operators transforming identically are
\begin{equation} \label{mix2}
 \frac{1}{3 \sqrt{2}} \left(
 {\cal O}^0_{\langle \langle 411 \rangle \rangle}
 - \half \left( {\cal O}^0_{\langle \langle 422 \rangle \rangle}
         + {\cal O}^0_{\langle \langle 433 \rangle \rangle} \right)
 \right)
\end{equation}
and
\begin{equation}
-\frac{1}{4 \sqrt{6}} \left(
 {\cal O}^5_{||123||} + 3 {\cal O}^5_{|123|} \right) \,.
\end{equation}
$\mbox{trace} = 0$, symmetric, $\tau^{(2)}_1$, $C=+1$:
\begin{equation} \begin{array}{l} \displaystyle
 {\cal O}^0_{\{1144\}} + {\cal O}^0_{\{2233\}} -
   {\cal O}^0_{\{1133\}} - {\cal O}^0_{\{2244\}}  \\
 \displaystyle \quad
  = \frac{1}{\sqrt{6}} \cdot \frac{\sqrt{3}}{\sqrt{2}} \left(
        {\cal O}^0_{\{1122\}} + {\cal O}^0_{\{3344\}} -
        {\cal O}^0_{\{1133\}} - {\cal O}^0_{\{2244\}} \right) \\
 \displaystyle \quad \hphantom{=} { }
  - \frac{1}{\sqrt{2}} \cdot \frac{1}{\sqrt{2}} \left(
        {\cal O}^0_{\{1122\}} + {\cal O}^0_{\{3344\}} +
        {\cal O}^0_{\{1133\}} + {\cal O}^0_{\{2244\}} -
      2 {\cal O}^0_{\{1144\}} - 2 {\cal O}^0_{\{2233\}} \right) \,.
\end{array}
\end{equation}
Operators transforming identically are
\begin{equation} \begin{array}{l} \displaystyle
 \frac{1}{6 \sqrt{2}} \Bigl(
 { }- {\cal O}^0_{1144} - {\cal O}^0_{4114}
    - {\cal O}^0_{1441} - {\cal O}^0_{4411}
    + 2 {\cal O}^0_{1414} + 2 {\cal O}^0_{4141}
\\  \displaystyle \hphantom{ \frac{1}{6 \sqrt{2}} \Bigl(}
 { }- {\cal O}^0_{2233} - {\cal O}^0_{3223}
    - {\cal O}^0_{2332} - {\cal O}^0_{3322}
    + 2 {\cal O}^0_{2323} + 2 {\cal O}^0_{3232}
\\  \displaystyle \hphantom{ \frac{1}{6 \sqrt{2}} \Bigl(}
 { }+ {\cal O}^0_{1133} + {\cal O}^0_{3113}
    + {\cal O}^0_{1331} + {\cal O}^0_{3311}
    - 2 {\cal O}^0_{1313} - 2 {\cal O}^0_{3131}
\\  \displaystyle \hphantom{ \frac{1}{6 \sqrt{2}} \Bigl(}
 { }+ {\cal O}^0_{2244} + {\cal O}^0_{4224}
    + {\cal O}^0_{2442} + {\cal O}^0_{4422}
    - 2 {\cal O}^0_{2424} - 2 {\cal O}^0_{4242} \Bigr)
\end{array}
\end{equation}
and
\begin{equation} \begin{array}{l} \displaystyle
 \frac{1}{2 \sqrt{6}} \Bigl(
      {\cal O}^5_{1234} - {\cal O}^5_{3214}
    - {\cal O}^5_{1432} + {\cal O}^5_{3412}
    + {\cal O}^5_{2143} - {\cal O}^5_{4123}
    - {\cal O}^5_{2341} + {\cal O}^5_{4321}
\\  \displaystyle \hphantom{ \frac{1}{2 \sqrt{6}} }
    + {\cal O}^5_{1243} - {\cal O}^5_{4213}
    - {\cal O}^5_{1342} + {\cal O}^5_{4312}
    + {\cal O}^5_{2134} - {\cal O}^5_{3124}
    - {\cal O}^5_{2431} + {\cal O}^5_{3421} \Bigr) \,.
\end{array}
\end{equation}
$\mbox{trace} = 0$, mixed symmetry, $\tau^{(8)}_1$, $C=+1$:
\begin{equation}
 2 {\cal O}^5_{2\{14\}} - {\cal O}^5_{1\{24\}}
                        - {\cal O}^5_{4\{12\}}
   = - \half {\cal O}^5_{\langle 124 \rangle }  \,.
\end{equation}
An operator transforming identically is
\begin{equation}
 \frac{\sqrt{3}}{6} \left( {\cal O}^0_{||311||}
                         - {\cal O}^0_{||344||}  \right) \,.
\end{equation}

\section{Discussion}

The results presented in this paper enable us to determine the mixings
allowed by symmetry among the lattice operators that are used to
calculate hadronic structure functions. Mixing with lower-dimensional
operators leads to contributions which diverge like a power of the
inverse lattice spacing in the continuum limit. These have either to
be subtracted nonperturbatively (a difficult task) or to be avoided by
a suitable choice of the operators. Mixing with operators of the same
dimension should be tractable by perturbation theory although a
nonperturbative treatment seems favourable. In any case, it leads to
additional uncertainties, but as $n$ grows, operators with no mixing
at all require more and more nonvanishing momentum components in the
calculation of their forward hadronic matrix elements, which makes
their Monte Carlo evaluation increasingly difficult. So some kind of
compromise is needed.

For the cases listed in section 6 a 1-loop calculation in
lattice perturbation theory leads to mixing only between the operators
(\ref{mix1}) and (\ref{mix2}). Moreover, the corresponding mixing
coefficient turns out to be rather small at the presently used values
of the coupling constant \cite{roma, letter}. Hence mixing with
operators of the same dimension seems to be relatively
harmless. However, the nasty mixing with lower-dimensional operators
will become unavoidable when dealing with operators of higher spin.

 \section*{Acknowledgements}
 This work is supported by the Deutsche Forschungsgemeinschaft and the
 European Community under contract number CHRX-CT92-0051.

\end{document}